\documentclass[8pt, preprint2]{aastex}
\usepackage[]{natbib}
\begin{document}

\title{An Abrupt Upper Envelope Cut-off in the Distribution of Angular Motions in Quasar Jets is compatible in all respects with a Simple Non-Relativistic Ejection Model}

\author{M.B. Bell\altaffilmark{1} and D.R. McDiarmid\altaffilmark{1}}

\altaffiltext{1}{Herzberg Institute of Astrophysics,
National Research Council of Canada, 100 Sussex Drive, Ottawa,
ON, Canada K1A 0R6;
morley.bell@nrc.gc.ca}

\begin{abstract}

A remarkable correlation is found in radio-loud quasars and BLLacs when the directly observed angular motions, $\mu$, of features ejected in the innermost regions of their jets are plotted on logarithmic scales versus the directly observed 15 GHz flux density, S, of their central engines: an abrupt upper envelope cut-off with a slope of 0.5 is obtained. This upper envelope and slope can be explained in a simple non-relativistic ejection model if (a), radio-loud quasars are radio standard candles and (b), for the sources defining the cut-off, the features are all ejected with similar speeds. The upper envelope is then due to the maximum projected velocity seen when the accretion disk is edge-on, and ejections are in the plane of the sky.
In our simple ejection model, where S is a good measure of relative distance, the observed distribution of angular motions can be explained if the radio luminosity of the source is a function of viewing angle, increasing towards face-on. In this scenario the flux densities of many of the sources with small viewing angles are increased above the detection limit, significantly altering the expected velocity distribution. This argument cannot be used in the cosmological redshift model, where Doppler boosting is then required.
Here we show that when $\mu$ is plotted versus redshift, $z$, the same upper envelope cut-off is seen. It is not as sharply defined, since, in this simple model, the $\mu$ upper envelope will be smeared out by sources lying at different cosmological distances, z$_{c}$. Normalizing all sources to the same distance (1 Jy) using the flux density, S, removes this smearing and improves the sharpness of the upper envelope, supporting our assumption that S is a measure of relative distance. In this model the redshift of quasars cannot be a reliable indication of their distance.

\end{abstract}

\keywords{galaxies: active - galaxies: distances and redshifts - galaxies: quasars: general}

\section{Introduction}

The motion of features ejected in the jets of quasars has been a fascinating topic for many years because of the need for relativistic ejection speeds to explain apparent superluminal motions in the cosmological redshift (CR) model. Bringing these objects closer removes the need for relativistic speeds, but then a large portion of the redshift must be intrinsic and no way of explaining large intrinsic redshifts has yet been found. As a result, the astronomical community has assumed for the last $\sim20$ years that ejection speeds are relativistic, and that Doppler boosting will have resulted in the preferential selection of sources with small jet viewing angles.

Here we discuss the abrupt upper cut-off in the angular motions $\mu$, seen on both $\mu$ vs $S$ and $\mu$ vs $z$ plots.

\section{A Simple, Non-Relativistic Ejection Model}

Although it was initially claimed that a simple ejection model could explain the observed distribution of ejection velocities in the CR model \citep{eke90}, this seemed to be ruled out when an analysis of the observed velocities showed that the probability that the sources had been picked at random from a parent population that is isotropically distributed was less than $10^{-5}$ \citep{coh90,kel04}. Because of this result it was concluded that the sources had to be preferentially aligned along the line of sight in order that the distribution of velocities could then be explained by Doppler boosting. Until now, no attempt has been made to explain the observed velocity distribution in a simple, non-relativistic ejection model in which a large portion of the observed redshift is intrinsic and \em the radio flux density is a good measure of relative distance. \em The observed velocity distribution can be easily explained in such a model, if the source radio radiation is anisotropic, increasing towards face-on orientations as the central engine becomes more visible through the hole in the central torus. (The sources approximate standard candles in respect of their radio flux density, S, at a $90\arcdeg$ viewing angle to the jet. S, for angles less than 90 deg is greater than this value, as just noted).  Sources with small viewing angles, $i$, relative to the line-of-sight, will then be shifted above the detection limit on log$\mu$ vs logS plots, changing the sin$i$ number distribution expected for number vs inclination plots to one in which the distribution can even decrease with increasing inclination angle. It is important to note that this is only true in the model proposed here in which S is a good measure of relative distance. Since this is not true in the CR model this argument cannot be used, and Doppler boosting must be invoked. 

In simple, non-relativistic ejection models, if active radio galaxies and quasars are ejecting material at reasonably similar velocities, the observed angular motion, $\mu$ (mas yr$^{-1}$), of these ejected features per unit time, will be proportional to the projected ejection speed, and will fall off inversely with distance for a given ejection angle relative to the line of sight. When plotted versus distance on logarithmic scales, a maximum value, or upper envelope, will be visible that will fall off with a slope of -1. This upper envelope is associated with ejection in the plane of the sky. i.e. $i$ = 90$\arcdeg$. For ejections closer to the line of sight, the observed angular motion will fall below this upper envelope.
Because the flux density, S, falls off inversely as the square of the distance, if the sources are radio standard candles, when $\mu$ is plotted versus the flux density, S, on logarithmic scales, the upper envelope, defined by sources with $i$ = $90\arcdeg$ will increase with a slope of 0.5. The angular motion then becomes a direct indication of distance, or a standard yardstick for a given viewing angle, and a measure of the projected velocity as the viewing angle changes. The observed flux density is thus dependent on both the distance of the source and its inclination angle $i$. If jet orientations are uniformly distributed over the sphere, the $i$-dependence will be sin$i$. However, it has been found that this is not the case, with many more sources than expected having small $\mu$-values \citep{bel06,kel04}. As noted above, this discrepancy can be explained if it is assumed, consistent with AGN models, that S is a function of $i$ such that it increases as $i$ decreases from 90$\arcdeg$ to 0$\arcdeg$. This dispersion is augmented if, as evidence presented below indicates, the maximum blob speed varies as (1+z)$^{-1}$, ie that this speed increases as z decreases. It is important to realize that the log$\mu \sim$ log S$^{0.5}$ relationship in the source frame for the solid line upper bound in Fig 1 is not compromised by either this speed dependence or by the anisotropy of source radiation.  Sources lying near the solid line in Fig 1 are oriented such that $i \sim90\arcdeg$ and many have a low value of z.

\section{Orientation Effects in AGNs}

Much work has been done in the last 10-20 years to try to come up with a unification scheme in which the observed properties of AGNs can be explained if the sources are all similar, but viewed at different orientations.
It is generally accepted today that orientation plays a large role when AGNs (radio galaxies, quasars, and BLLacs) are studied \citep{ant93,hoo97,kee80,nan97,pog89,sch01}, and \citet{urr95} have discussed some of the consequences that are introduced when obscuration in the central region of AGNs results in an anisotropic radiation pattern.



\begin{figure}
\hspace{-1.0cm}
\vspace{-1.0cm}
\epsscale{1.0}
\plotone{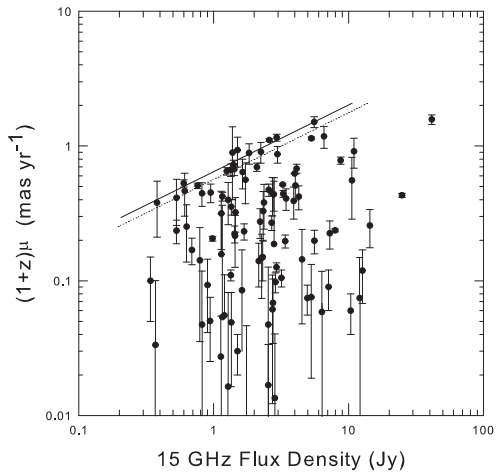}
\caption{{Angular motion in the source frame for 96 quasars and BLLacs plotted against maximum observed flux density from \citet{kel04}. See text for an explanation of the dotted line. \label{fig1}}}
\end{figure}

\begin{figure}
\hspace{-1.0cm}
\vspace{-1.2cm}
\epsscale{0.9}
\plotone{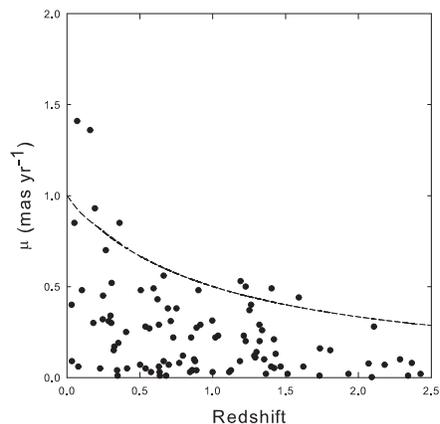}
\caption{{Angular motion, $\mu$, plotted vs redshift. The dashed line shows how the angular motion is expected to fall off with redshift if redshift is largely intrinsic and time dilation plays a role. \label{fig2}}}
\end{figure}

\section{Angular Motion Data}

A large sample of radio-loud quasars with active jets has been monitored for the last $\sim10$ years using the VLBA, in order to follow the motions of the ejected \em blobs \em in their inner jets \citep{kel04}. This is an excellent set of data and it was recently used \citep{bel06} to carry out an extensive investigation of the model proposed above. In Fig 1, the maximum angular motion per unit time in the source reference frame, $\mu$(1+z), observed for 96 $superluminal$ quasars and BL Lac objects from \citet{kel04} is plotted versus the 15 GHz VLBA flux density, S, of the central engine, on logarithmic scales. Here, S is the maximum, total flux density from the core and innermost jet, observed over several years. When more than one $\mu$ has been observed in a source, the highest value has been used. The 13 sources with large motion uncertainties have not been included.
These data are plotted without the error bars in Fig 1 of \citet{bel06}.
The solid line along the upper envelope in Fig 1 has a slope of 0.5 and is a good fit to the data, as predicted in the above simple model. This is a remarkable result, and is evidence in support of the non-relativistic ejection model proposed here. The sharpness of the upper envelope seen in Fig 1 requires that both the ejection velocities and the maximum output power achieved be similar in all sources, and may imply that something sets an upper limit to these quantities, perhaps similar to the maximum value attained in the power peak seen in SnIa explosions. For this reason the maximum flux density value was used.
The radio galaxies were omitted because of the chance that there may be confusion in these objects between the core component and the radio flux density originating in the surrounding host galaxy. Alternatively the core component might even be reduced by the presence of the host galaxy (but see below). Most importantly, the fact that the upper envelope is well defined indicates that the flux density variations have not had a significant affect, which may be due to the fact that, as for SnIae, only the peak values were used.

For clarity we now itemize the relevant model assumptions we use to explain Fig 1. They are:

(a) The radio flux density S is a good measure of relative distance (sources are radio standard candles) for a given viewing angle.

(b) The ejection velocity of the fastest observed ejections in each of these sources are similar, or at least have a maximum value. 

(c) Jet orientations are random and radio emission is anisotropic increasing towards face on.

With these assumptions, the slope of 0.5 is explained by assumptions (a) and (b), its sharpness is explained by assumptions (a) and b) together, and the vertical distribution is explained by (c) but only if (a) is true. This model can thus explain Fig 1 without Doppler boosting, unlike the CR model where (a) is not true.

 
\begin{figure}
\hspace{-1.0cm}
\vspace{-1.4cm}
\epsscale{0.9}
\plotone{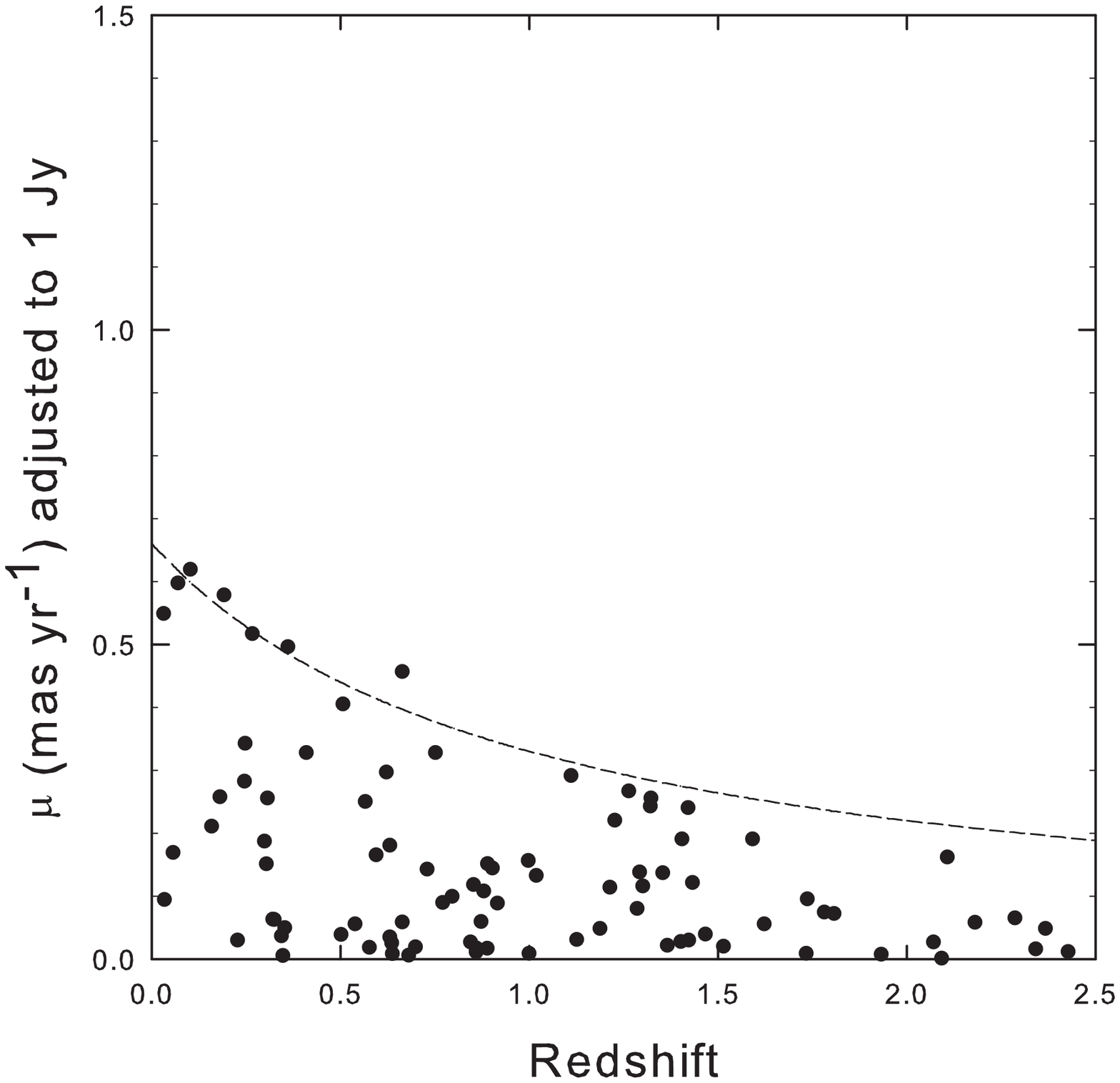}
\caption{{Angular motion, $\mu$, corrected to 1 Jy distance, plotted vs redshift. The dashed line shows how the angular motion is expected to fall off with redshift if redshift is largely intrinsic and produces time dilation. \label{fig3}}}

\end{figure}

In the simple, non-relativistic ejection model considered here, the sources that lie along the upper envelope in Fig 1 will have their jets directed at close to 90$\arcdeg$ to the line of sight. Those that are located near the bottom of the plot will have their jets directed closer to the line of sight ($i < 30\arcdeg$).  If the jets are oriented in completely random directions, and no selection effects are involved, the distribution of sources in Fig 1 would be expected to be concentrated much nearer to the upper cut-off than is found, as pointed out previously \citep{kel04,coh90,bel06}. In fact, the dotted line in Fig 1 indicates the line above which 50 percent of the sources should be located. Clearly, if $\mu$ is a measure of the inclination angle of the jet, many more sources than expected have orientations below this line, and sources with inclinations near $90\arcdeg$ are much rarer than the sin$i$ distribution for random orientations would predict. 
As noted above, this apparent discrepancy is easily explained in the model proposed here, where the flux density is a measure of distance, if the radiation from the central engine is directed, with the observed flux density increasing from edge-on to face-on, as the region inside the hole in the torus becomes visible.
Since most of the activity in the central engine is likely to be located in the vicinity of the accretion disk surrounded by an opaque torus, an orientation effect is not an unlikely scenario. In fact, it is the very principle upon which the Unification Model for AGNs is based. This means that in the simple model considered here, the flux density can only be a good standard candle, and a measure of relative distance, for a fixed inclination angle. However, this will not affect the slope of the upper envelope in Fig 1 since the sources that lie along this sharp cut-off all have the same edge-on orientation.

We are currently looking into whether or not it may be possible to estimate the opening angle of the torus using these data. Preliminary results indicate that when a more rigorous analysis is used the opening angle is much larger than was suggested previously by \citet{bel06}.

\section{Angular Motion versus Redshift}

The simple, non-relativistic ejection model discussed here, in which $\mu$ represents the projected velocities of the $blobs$, naturally produces an upper limit to the $\mu$ vs S plot with a slope of 0.5, if S is a measure of relative distance. In this model, if the true distance to the sources ($z_{c}$) is small, from the relation (1+z) = (1+$z_{i}$)(1+$z_{c}$), the measured redshift ($z$) must then be mostly intrinsic ($z_{i}$), and the $\mu$ vs $z$ plot will also have an upper envelope cut-off. This may not be as clearly defined, since at each intrinsic redshift value there will be sources that lie at different cosmological redshifts ($z_{c}$). Although this distance component is expected to be small, since $\mu$ is proportional to distance it will tend to smear out a sharply defined upper envelope in a $\mu$ vs $z$ plot

The $\mu$ vs $z$ plot is also worth examining for a second reason. It can also give information on the nature of the intrinsic redshift. In non-standard redshift models, such as that proposed here, where a large component of the quasar redshift is not related to the expansion of the Universe, it has been proposed that quasars are born with a large intrinsic redshift component, out of the nuclei of mature, active galaxies \citep{arp97,arp98,arp99,bel02a,bel02b,bel02c,bel04,bel06,bel06a,bur04,bur99,gal05,nar80,nar93}, and that this intrinsic component decreases as the object matures. In these models the redshift of all the matter in the object depends on its age. The entire source is then expected to have the same redshift. In such a model, how the upper envelope of the $\mu$ vs $z$ plot varies with redshift will depend on whether or not time dilation is active and whether or not the ejection velocity varies with intrinsic redshift. As the sources evolve into mature galaxies, and become more luminous in this model, one might expect that the velocity at which the features are ejected would increase. It was not possible to obtain this information from Fig 1 because the high- and low-redshift sources are scattered throughout the plot.

In Fig 2, which again represents the quasars and BLLac objects taken from \citet{kel04}, $\mu$ is plotted vs redshift, $z$. An upper envelope cut-off is again visible, as predicted above, and the dashed line shows how the motion along this envelope is expected to fall off with increasing redshift due to a time dilation factor of 1/(1+z). The upper envelope follows this curve closely, indicating not only that there is an upper cut-off, as is predicted for a maximum velocity corresponding to ejection in the plane of the sky in our model, but also that, if the redshift is intrinsic, it may produce a time dilation effect like that produced by other known redshifts. This result, together with the result obtained in Fig 1, are both evidence in support of the conclusions that the upper envelope represents a natural upper limit to the angular motion that corresponds to a maximum projected ejection velocity seen for ejection in the plane of the sky. These results are then completely compatible with the decreasing intrinsic redshift model referenced above, which assumes that the redshifts of quasars are mostly intrinsic.


The angular velocity vs redshift plot was examined previously by \citet{coh88} who argued that the upper bound to this plot could be explained in the CR model. From the data available at that time it was argued that the upper envelope corresponded to a maximum value of $\beta_{app}$ that corresponded to a Lorentz factor of $\gamma$ = 13. Using the present source sample \citet[see their Fig 11]{kel04} found that the upper envelope in the angular velocity-redshift plot corresponded to a maximum value of $\beta_{app}$ of $\gamma$ = 25, although the fit was admittedly crude. One can then argue from these two results that, for any given set of data, one can always find a gamma which corresponds to an upper bound. Confidence is not produced, however, when the gammas differ significantly from one dataset to another. Several other attempts have been made to explain this upper envelope in the CR model, but they are far from convincing. Most contain arguments that require elaborate modeling (see for example \citet{coh88,ver94,lis97}), using the many CR model variables available when relativistic velocities are assumed. In the simple, non-relativistic model discussed here, the upper envelopes in Figs 1 and 2 are immediately explained by the maximum projected velocity associated with edge-on orientations. Furthermore, $\mu$-values in this model are not expected to cluster near the maximum value (upper envelope) in Figs 1 or 2, if the source radiation is a function of orientation, increasing towards face-on.

\begin{figure}
\hspace{-1.0cm}
\vspace{-1.6cm}
\epsscale{1.0}
\plotone{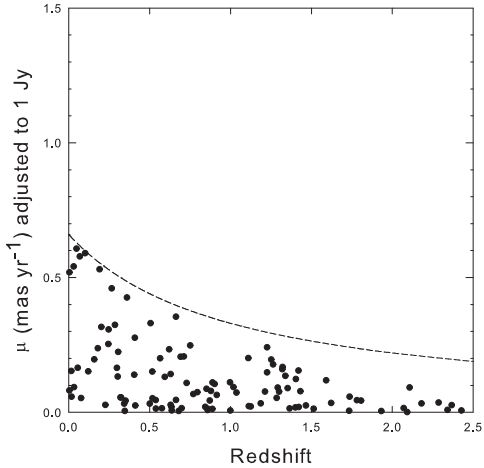}
\caption{{Fig 3 after applying a k-correction of (1+z) to S before adjusting to 1 Jy. The dashed line shows how the upper envelope is expected to fall off due to time dilation.\label{fig4}}}
\end{figure}
 
\begin{figure}
\hspace{-1.0cm}
\vspace{0.0cm}
\epsscale{0.9}
\plotone{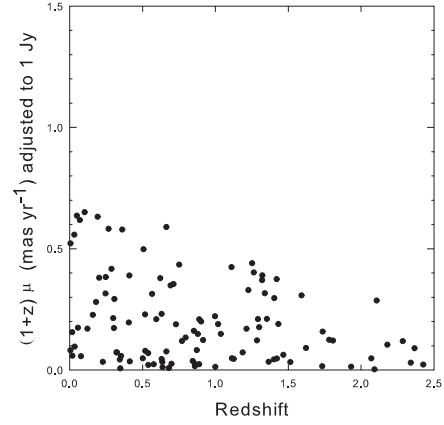}
\caption{{Fig 4 after a correction of (1+z) for time dilation. \label{fig5}}}
\end{figure}
 
\begin{figure}
\hspace{-1.0cm}
\vspace{-1.6cm}
\epsscale{1.0}
\plotone{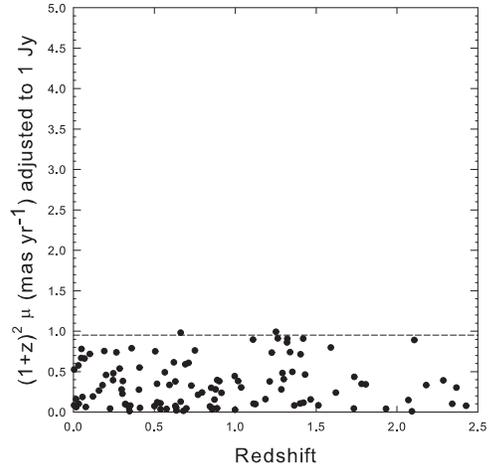}
\caption{{Fig 5 after $\mu$ corrected by (1+z) to account for $\mu$ decrease with redshift. The latter is assumed here to be related to the significant decrease in source luminosity with increasing intrinsic redshift. \label{fig6}}}
\end{figure}

\begin{figure}
\hspace{-1.0cm}
\vspace{-1.0cm}
\epsscale{1.0}
\plotone{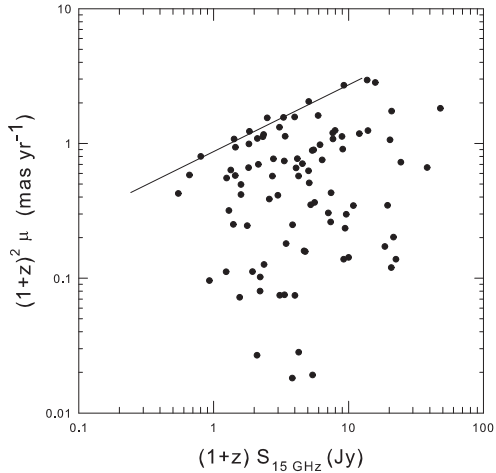}
\caption{{Fig 1 after k-correction of (1+z) to S, with $\mu$ corrected by a factor of (1+z) for time dilation and (1+z) for $\mu$-change with $z$. \label{fig7}}}
\end{figure}

\begin{figure}
\hspace{-1.0cm}
\vspace{-1.0cm}
\epsscale{0.9}
\plotone{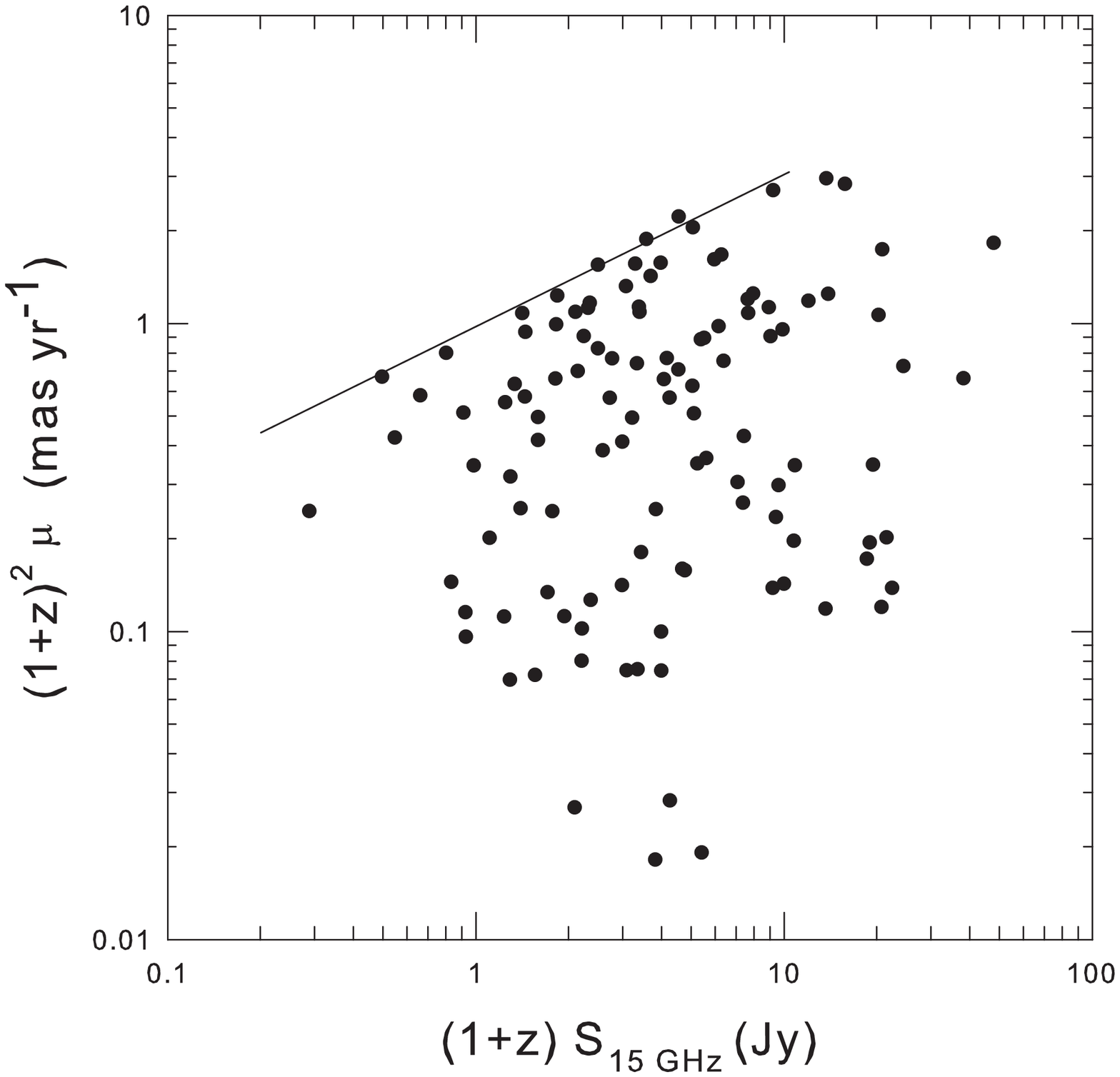}
\caption{{Same as Fig 7 with radio galaxies included. See text for a description of the sources plotted as open characters. \label{fig8}}}
\end{figure}




Although an upper envelope cut-off is clearly visible in Fig 2, and its slope is close to what is expected if time dilation effects are present, there is a further test that can be undertaken using this plot to confirm our claim that the flux density is a measure of relative distance. If, as proposed here, the redshifts in Fig 2 are mostly intrinsic, as noted above the sources at each intrinsic redshift value can cover a range of distances (range of $\mu$-values) that will tend to smear out the upper cut-off. If S is truly a measure of relative distance, it can then be used to normalize all sources to the same distance, which should remove the smearing effect and produce a much sharper upper cut-off. This, of course, will only work if S is truly a measure of relative distance. When $\mu$ is adjusted to a distance corresponding to 1 Jy using the relation $\mu$(1Jy) $\propto \mu$S$^{-1/2}$ and replotted in Fig 3 it is obvious that the upper envelope is now much more clearly defined over the entire range of redshifts. This works because $i \sim90\arcdeg$ for the sources near the upper bound.  \em This result then supports our original assumption that the flux density is a good measure of relative distance for radio loud quasars and BLLacs. \em Although there is no obvious reason why there should be an abrupt upper envelope cut-off in the CR case, the upper envelope predicted for $\gamma$ = 25 in \citet[see their Fig 11]{kel04} is clearly not as good a fit to the data as is the fit in Fig 3.

\section{K-corrections to the Flux Density}

In Figs 1-3 above, the k-correction was assumed to be small since these sources were selected because they have flat spectra. However, this needs to be examined in more detail. Since the k-correction factor is proportional to $(1+z)^{1+\alpha}$ for S varying as $\nu^{-\alpha}$, then even if the spectral index $\alpha$ = 0, a correction factor of (1+z) still needs to be applied to the flux density in Figs 1, 2, and 3. Fig 3 has been re-plotted in Fig 4 after applying a k-correction of (1+z) to the flux densities before normalizing the $\mu$-values to a distance corresponding to 1 Jy. 
Although an upper envelope cut-off is still clearly defined, it now drops off with redshift more quickly than before the k-correction was applied. In Fig 5 a time dilation correction factor of (1+z) has been applied to the data in Fig 4. There is still evidence that the upper envelope falls off with increasing redshift. In the DIR model the optical host is significantly less luminous at higher intrinsic redshifts. If this decrease in luminosity is associated with a lower ejection speed, this can explain the slight upper envelope fall-off in Fig 5. Although the fall-off is roughly proportional to 1/(1+z), as is demonstrated in Fig 6, where the $\mu$-values in Fig 5 have been multiplied by a second factor of (1+z) before plotting, the true correction factor may be more complicated. Although there may be a decrease in ejection speed with increasing redshift, it is worth noting that this decrease is quite small, changing only by a factor of 3 between $z$ = 0 and $z$ = 2.

Applying a k-correction to the flux densities, and a factor of (1+z) to correct for the change in $\mu$ with redshift before $\mu$(1+z) is plotted versus S, shows clearly in Fig 7 that the upper envelope in Fig 1 is still sharply defined and has the required slope of 0.5 predicted by the proposed model. It is also worth noting that we have identified the sources that define the upper envelope in Fig 7 and they are found to be the same ones that define the upper envelope in Figs 4, 5, and 6.

\subsection{Radio Galaxies as Standard Candles}

It was assumed above that as the quasars matured into BLLacs, and eventually to radio Galaxies, that there might be a component of the flux density originating in the surrounding host galaxy that would add to the flux density of the core, destroying its standard candle properties. For this reason radio galaxies were not initially included in the sample. In Fig 8 we have included the 24 radio galaxies and find that they too appear to be equally good standard candles when a k-correction is applied, and adjustments are made to account for an apparent increase of a factor of $\sim3$ in the ejection velocity with decreasing redshift. The corrections applied to the data, since they are redshift based, do not significantly alter the $\mu$ and S values of the radio galaxies whose redshifts are all quite small. If the flux densities of these sources are then still good standard candles, it implies that the 15 GHz VLBA flux density originates mainly in the core and inner jet region, and not in the surrounding galactic component, although this may not be too surprising if any broad, galactic component is resolved out.


\section{Conclusions}

Empirical evidence has been presented using the directly observed parameters $\mu$, S, and $z$, that allows the following conclusions to be drawn for radio-loud quasars, BLLacs and radio galaxies:

1) An upper envelope with a slope of 0.5 is seen when $\mu$ is plotted versus S on logarithmic scales, that is easily explained in a very simple ejection model in which (a) the ejection velocities are similar but non-relativistic, (b) the sources are radio standard candles whose radiation is anisotropic with a radio brightness that is a function of viewing angle similar to that seen in related AGNs, and (c) the angular motions, $\mu$, of the features in their inner jets are a measure of their projected ejection velocity (and therefore proportional to viewing angle).

2) In the non-relativistic model proposed here, if $\mu$ is a measure of projected ejection velocity, the range of $\mu$-values measured (from $\mu$ = 0.05 to $\mu$ = 1), corresponds to a range of viewing angles from a few degrees to $90\arcdeg$.

3) The upper envelope cut-off, produced naturally by our simple, non-relativistic ejection model, immediately explains the upper envelopes seen in \citet[figs 6 and 11]{kel04}.


4) When $\mu$ is plotted versus redshift an upper envelope cut-off is also obtained. It is easily explained in our simple ejection model, but is again difficult to explain in the CR model. Moreover, the sharpness of this cut-off is improved when $\mu$-values are adjusted to the same cosmological distance using the flux density as an indication of relative distance, which further supports our simple, non-relativistic model. After time dilation is taken into account, the upper envelope falls off smoothly by a factor of $\sim3$ between redshifts of 0 and 2.5, which can be explained in the simple ejection model by a slight increase in the ejection speed as the objects become more luminous and turn into mature galaxies.




These results suggest that the model proposed earlier \citep{eke90,kel72} may have been correct after all,
and the evidence presented here does imply that intrinsic redshifts are a real possibility. No matter how slight the reader feels this possibility may be, it needs to be given a fair examination.


\end{document}